\documentclass[twocolumn,showpacs,aps,prl,preprintnumbers,amsmath, amssymb,superscriptaddress]{revtex4}

\usepackage{graphicx}
\usepackage{dcolumn}
\usepackage{bm}

\begin{document}


\title{High-$Q$ impurity photon states bounded by a photonic-band-pseudogap \\in an optically-thick photonic-crystal slab}


\author{Se-Heon Kim}
\email{seheon@caltech.edu}
\affiliation{Department of Physics, California Institute of Technology, Pasadena, CA 91125, USA}
\affiliation{Kavli Nanoscience Institute, California Institute of Technology, Pasadena, CA 91125, USA}

\author{Andrew Homyk}
\affiliation{Kavli Nanoscience Institute, California Institute of Technology, Pasadena, CA 91125, USA}
\affiliation{Department of Electrical Engineering, California Institute of Technology, Pasadena, CA 91125, USA}

\author{Sameer Walavalkar}
\affiliation{Kavli Nanoscience Institute, California Institute of Technology, Pasadena, CA 91125, USA}
\affiliation{Department of Electrical Engineering, California Institute of Technology, Pasadena, CA 91125, USA}

\author{Axel Scherer}
\affiliation{Department of Physics, California Institute of Technology, Pasadena, CA 91125, USA}
\affiliation{Kavli Nanoscience Institute, California Institute of Technology, Pasadena, CA 91125, USA}
\affiliation{Department of Electrical Engineering, California Institute of Technology, Pasadena, CA 91125, USA}



\date{\today}

\begin{abstract}
We show that, taking a two-dimensional photonic-crystal slab system as an example, surprisingly high quality factors ($Q$) over $10^5$ are achievable, even in the {\em absence} of a rigorous photonic-band-gap. We find that the density of in-plane Bloch modes can be controlled by creating additional photon feedback from a finite-size photonic-crystal boundary that serves as a {\em low}-Q resonator. This mechanism enables significant reduction in the coupling strength between the bound state and the extended Bloch modes by more than a factor of 40.
\end{abstract}



\maketitle


Significant reduction in the radiation rate of a point-like emitter can be achieved by setting up mirrors around it,\cite{Purcell46} or by employing photonic crystals (PhCs),\cite{Yablonovitch87,Sajeev87} which is a photonic analogue of atomic crystals for electron waves.\cite{Joannopoulos_book} It has long been believed that the existence of the photonic-band-gap (PBG) is essential to achieving a {\em spatially localized} high-$Q$ electromagnetic bound state using a PhC cavity. Thus, most efforts so far have focused on artificial dielectric structures possessing a PBG.\cite{Yablonovitch_93, Noda00, Chow_Nature_00, Maldovan_NatMat} Donor- or acceptor-like impurity photon states can be formed at the location of a crystal defect.\cite{YablonvitchGm91}  Such a localized state (with small mode volume, $V$) has drawn much attention in the context of cavity quantum electrodynamics (cQED) experiments\cite{Yoshie_Nature_04, Noda_Nat_Photon_07} where the use of high $Q/V$ cavities are essential to enhancing light-matter interaction. 

\begin{figure*}[t]
\centering
\includegraphics[width=13.5cm]{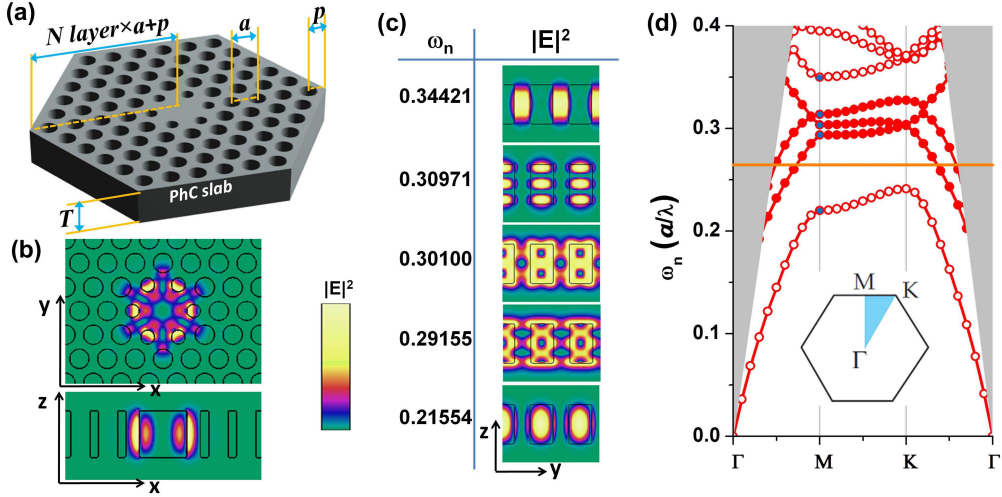}
\caption{\label{fig:fig1} (a) A 3-D rendering of the photonic-crystal cavity to be studied in this Letter. The number of photonic-crystal hole layers ($N$ $layer$) and the boundary termination ($p$) are to be varied. $a$ is the lattice constant of the 2-D triangular-lattice photonic-crystal with the background air-holes, $R$, of $0.35 a$. The refractive index of the slab is assumed to be 3.4. The six nearest neighbor holes are reduced ($Rm = 0.25 a$) and slightly pushed away (by a scale factor of 1.1) from the defect center. (b) $|{\bf E}|^2$ distribution of the hexapole mode in a $T$ = $1.731 a$ slab. (c) Cross-sectional views of the first five lowest Bloch modes at $M$-point. These modes all have even symmetries with respect to the plane at $z$ = 0. (d) The photonic band structure for all even Bloch modes. The hexapole mode resonance shown in (b) ($\omega_n \sim 0.265$) is represented by the horizontal line, in which $\omega_n$ is the normalized frequency defined by $\omega_n \equiv a / \lambda$.}
\end{figure*}

Due to fabrication related difficulties, three-dimensional (3-D) PhCs have been replaced with a lower dimensional counterpart relying on index guiding in one or two dimensions. Often, this assumes the from of a thin dielectric slab\cite{JohnsonFa99}, whose thickness ($T$) is roughly equal to half the effective wavelength ($T \approx \lambda / 2n_{eff}$) in order to optimize the size of the in-plane PBG. One representative such a design is shown in Fig.~\ref{fig:fig1}(a) and (b). Note that a 2-D PhC with the triangular-lattice of air-holes supports the PBG only for {\em even} guided modes, where the symmetry is defined with respect to the mirror plane at $z = 0$. However, this {\em incomplete} PBG does not preclude the possibility of high-$Q$ defect states, because the same mirror symmetry ensures that the defect mode with the {\em even} symmetry will be {\em completely} decoupled from all {\em odd} guided modes. Lowering the dimensionality often creates new symmetries that can be exploited, making the condition for localized photon states less stringent. 


It is well established that an optically-{\em thick} PhC slab does not support any PBG for both {\em even} and {\em odd} symmetries\cite{JohnsonFa99, SHKim_OL2012}. As an example, in Fig.~\ref{fig:fig1}(d), we present a photonic band structure ($\omega$-$k$ diagram) for {\em even} Bloch modes in a {\em thick} PhC slab with $T = 1.731 a$. The formation of the PBG is hindered by the higher-order slab modes lying between the 1st and the 5th bands. Note that the energy gap defined by the 1st and the 5th bands ($\triangle \omega_{1-5}$) will remain more or less constant as $T$ increases and, in the limit of $T \rightarrow \infty$, $\triangle \omega_{1-5}$ will approach the PBG of an ideal 2-D PhC\cite{JohnsonFa99}. Also note that, though all bands, ${\bf E}_{(Bloch)} ({\bf r})$, shown in Fig.~\ref{fig:fig1}(d) are mutually orthogonal, the bound state in a defect region, ${\bf E}_{(cav)} ({\bf r})$, in general, {\em can} couple to any of the higher-order Bloch modes. However, this coupling strength, $|\kappa_{c,B}| \sim |\int d^3{\bf r}~\triangle \epsilon {\bf E}_{(cav)} \cdot {\bf E}^{*}_{(Bloch)} |$\cite{Haus_book}, shouldn't be strong, because ${\bf E}_{(cav)} (z)$ resembles ${\bf E}_{(Bloch)} (z)$ of the 1st band. In this sense, any optically-thick PhC slab bears a {\em pseudo}-PBG. Similar pseudo-energy gap in solid state physics can create {\em resonance} states\cite{Madelung_book}. Here, the photonic counterpart can be easily manipulated with high precision by means of the mature modern nanofabrication technology.


In this Letter, we will show how  $\kappa_{c,B}$ can be controlled by setting up simple mirror boundaries around the finite size PhC resonator [Fig.~\ref{fig:fig1}(a)]. The additional boundary conditions imposed by the mirrors can alter the density of Bloch states in the momentum space (${\bf k}$-space). Thus, in relation to the ${\bf k}$ spectrum of ${\bf E}_{(cav)}$\cite{Srinivasan_02}, $\kappa_{c,B}$ can experience significant change. We note that the situation is analogous to the well-known cQED example of a point dipole source ($\approx$ the hexapole mode) in an optical resonator ($\approx$ the low-$Q$ resonator defined by the boundary termination)\cite{Hinds}.


We perform fully-vectorial 3-D numerical simulations using the finite-difference time-domain method (FDTD) to understand the mutual interaction between ${\bf E}_{(cav)}$ and ${\bf E}_{(Bloch)}$. First, we study the energy decay rate ($\gamma$) of the hexapole mode\cite{H_Y_Ryu_03} shown in Fig.~\ref{fig:fig1}(b). The total decay rate ($\gamma_{tot}$) can be decomposed into the decay rates into the horizontal direction ($\gamma_{horz}$) and the out-of-plane direction ($\gamma_{vert}$). Then, $\gamma$ is translated into $Q$ factor through $Q \equiv \omega / \gamma$. Thus, $1/Q_{\rm tot} = 1/Q_{\rm horz} + 1/Q_{\rm vert}$\cite{SHKim_OL2012}. Note that we set up detection planes for the Poynting energy flux ($\sim {\bf E} \times {\bf H}^*$) away from the mirror boundary, so that $\gamma_{horz}$ accounts for the reflection/transmission at the PhC-air discontinuity.  
 
\begin{figure}[b]
\centering\includegraphics[width=8.5cm]{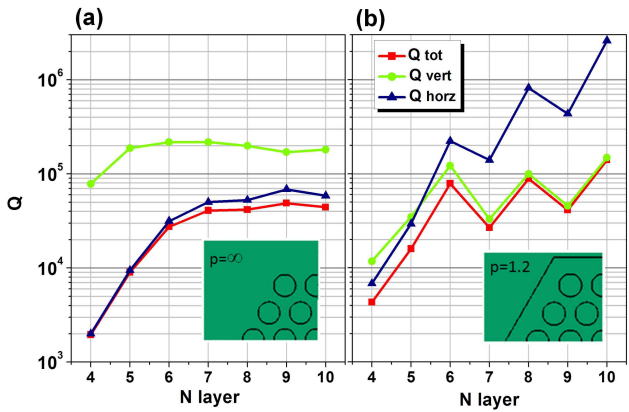}
\caption{\label{fig:fig2} $Q_{tot}$, $Q_{vert}$, and $Q_{horz}$ of the hexapole mode for two different boundary terminations, (a) $p = \infty$ and (b) $p = 1.2 a$. When the mirror boundary effect is {\em weak} ($p = \infty$), $Q_{horz}$ is limited by about 60,000 due to the coupling to the in-plane Bloch modes. However, in the case of the abrupt termination by air, $Q_{horz}$ can be made over 2,600,000.}
\end{figure}

In Fig.~\ref{fig:fig2}, we consider the two boundary terminations of $p = \infty$ and $p = 1.2 a$. For $p = \infty$, $Q_{tot}$ tends to be saturated by $Q_{horz}$; $Q_{tot}$ approaches $44,000 \sim 48,000$ as $N$ layer $>$ 8\cite{SHKim_OL2012}. In fact, this particular choice of the boundary termination $p = \infty$ approximately ensures negligibly small reflection off the interface. Alternatively, we can simulate the transparent boundary condition by overlapping the perfectly matched layer in the FDTD with the PhC air-holes\cite{PML_PhC}, which results in the similar saturated $Q_{horz}$ of $\sim$60,000 ($\equiv Q^{(sat)}_{horz}$).

\begin{figure*}[t]
\centering\includegraphics[width=16cm]{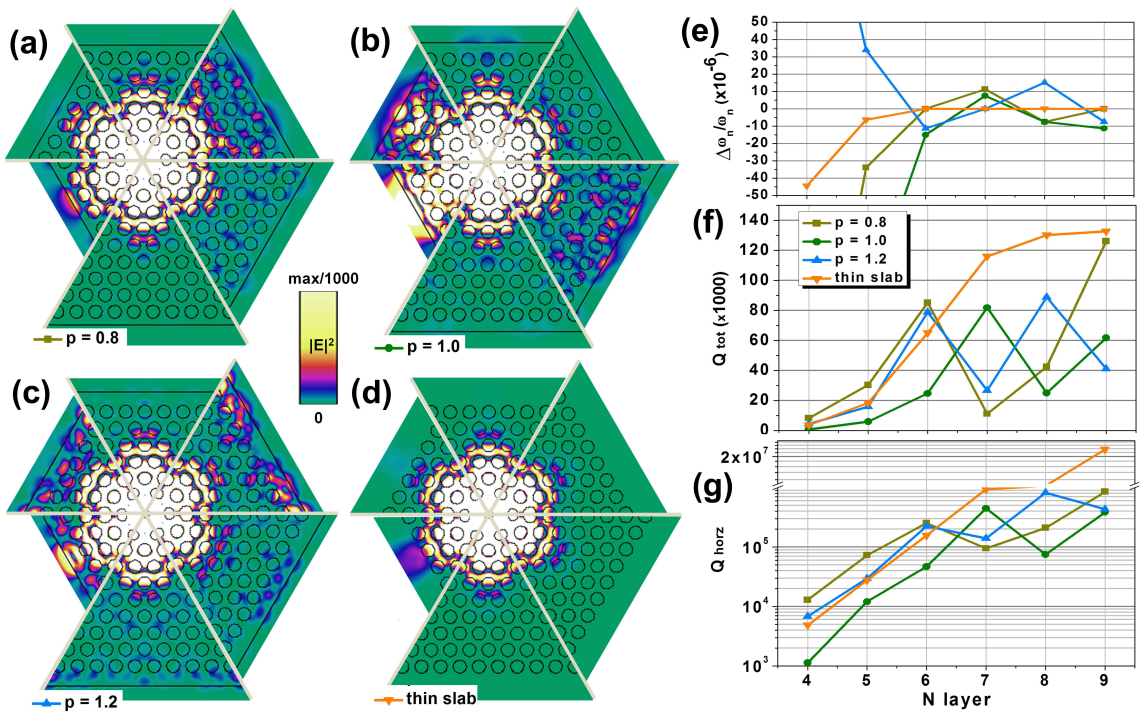}
\caption{\label{fig:fig3} (a-c) Top-down views ($|{\bf E}|^2$) of the hexapole resonances in a $T = 1.731 a$ slab as we increase $N$ $layer$ from 4 to 9 (in the clockwise direction) for three different boundary terminations of $p = 0.8, 1.0$, and $1.2 a$, respectively. (d) The hexapole resonances in a {\em thin} ($T = 0.488 a$) slab for $p = \infty$ are presented in the same manner used in (a-c). Note that this thin slab possesses the in-plane photonic-band-gap. (e) $Q_{horz}$ and (f) $Q_{tot}$ for the hexapole resonances shown in (a-d).}
\end{figure*}

Usually, the reflection coefficient at the PhC-air boundary is not so large as 0.5$\sim$0.6\cite{PhC_mirror_R}. However, drastically different behaviors can be seen by forming simple mirrors with $p = 1.2 a$ [Fig.~\ref{fig:fig2}(b)]. (i) All $Q$ values {\em strongly} modulate with $N$ $layer$. (ii) $Q_{horz}$ can be brought up to a surprisingly high value of $\sim 2.6 \times 10^6$ at $N$ $layer$ = 10. This implies that $\gamma_{horz}$ can be reduced by more than a factor of 40 compared with the case of minimal reflection ($p = \infty$). (iii) Since $Q_{horz} >> Q_{vert}$, now the hexapole mode emits more photons into the out-of-plane direction. Interestingly, all $Q$s modulate in the same fashion; whenever $Q_{horz}$ is peaked, so is $Q_{vert}$. Therefore, the abrupt boundary termination does not simply redirect the in-plane guided energy into the out-of-plane radiation. Rather, there should be a common physical principle for the observed enhancement and suppression. Moreover, note that the maximum $Q_{tot}$ of $1.4\times10^5$ in the present case can be further increased by employing a higher $Q_{vert}$ cavity mode. Also note that $Q_{vert}$ of the hexapole mode is slightly lowered by the use of finite space grids ($\triangle = a/20$) used in the FDTD simulation\cite{Kim_PRB_06}. 


More detailed analyses are performed as we finely tune the mirror boundary conditions. Specifically, we tune $p$ in the range of $0.8 - 1.2 a$ as a means to control the phase shift ($\phi$) upon reflection. Here, we will provide direct {\em graphical} evidence for the enhancement/suppression of the in-plane Bloch modes. In order to visualize the very weak near-fields in the outskirt regions, we adopt the saturated color scheme in which $1/1000$ of the intensity maximum is taken as the upper bound. In Fig.~\ref{fig:fig3}(a)-(c), we present $|{\bf E}_{(cav)} ({\bf r}_{||}, z=0)|^2$ of the hexapole mode for three different $p = 0.8$, $1.0$, and $1.2 a$. We take only one-sixth area (cut by the two $\Gamma$-$K$ lines) of the hexapole mode profile for each $N$ $layer$, then combined with plots of different $N$ $layer$s to create a one image filling upto 360$^{\circ}$. We also provide a similar plot for a PBG-confined hexapole mode in Fig.~\ref{fig:fig3}(d), corresponding to a slab thickness of $0.488 a$.  For quantitative analyses, we also provide graphs for $Q_{tot}$ [Fig.~\ref{fig:fig3}(f)] and $Q_{horz}$ [Fig.~\ref{fig:fig3}(g)] as well as the resonance frequency shift $\triangle \omega_n / \omega_n$ [Fig.~\ref{fig:fig3}(e)].

We find that, at ($N$ $layer$, $p$) = (9, $0.8 a$), $|{\bf E}_{(Bloch)}({\bf r}_{||})|^2$ is {\em almost completely suppressed}, which is comparable with that of the PBG-confined case shown in Fig.~\ref{fig:fig3}(d). On the other hand, at certain combinations of ($N$ $layer$, $p$), $|{\bf E}_{(Bloch)}({\bf r}_{||})|^2$ becomes quite strongly pronounced; for example, see (7, $0.8 a$), (8, $1.0 a$), and (7, $1.2 a$). For small $N$ $layer$ such as at ($N$ $layer$ $\leq 5$, $p = 1.0 a$), it seems natural to have more intense photon tunneling than the equivalent PBG-confined case. However, by {\em slightly} tuning $p$, the strong in-plane loss in the previous case can be greatly reduced by more than a factor of 10, whose resultant $Q_{tot}$ and $Q_{horz}$ can be higher than those of the PBG-confined case\cite{thin_slab_Q}. This result may find a practical importance in applications where a device miniaturization is required while keeping $Q$ reasonably high. 

We find that $|{\bf E}_{(Bloch)}({\bf r}_{||})|^2$ variations can explain peaks and dips observed in $Q_{horz}$ or $Q_{tot}$. We also note that the more intensified $|{\bf E}_{(Bloch)}({\bf r}_{||})|^2$ in the outskirt region can contribute to the excessive scattering losses into the vertical direction, thereby lowering $Q_{vert}$. In short, what we have shown here is that $\kappa_{c,B}$ strongly depends on the detailed boundary conditions. $\kappa_{c,B}$ obviously depends on the size of the hexagonal boundary, which determines the phase thickness of the 2-D cavity. It is also understandable that $p$ is a very critical parameter controlling the density of in-plane Bloch modes, as has been seen in many cQED examples of an atom and a cavity. Further evidence of this analogy can be found in Fig.~\ref{fig:fig3}(e), which reports the fractional frequency shift. Even using sufficiently thick PhC mirrors with $N$ $layer \geq 6$, the thick-slab cases show noticeable modulations in $\triangle \omega_n / \omega_n$ order of $\pm 10^{-5}$, while the PBG-confined case does not. These energy-level shifts\cite{Hinds} are signatures of the coupling between ${\bf E}_{(cav)}({\bf r})$ and ${\bf E}_{(Bloch)}({\bf r})$ and can be explained in terms of $\kappa_{c,B}$\cite{Haus_book}.


Though the FDTD provides very accurate first-principle means to understand the $\kappa_{c,B}$ modulation, we have not been quite convinced as to how the simple boundary termination (hence the {\em low}-$Q$) can enable such large modulations in $Q_{horz}$ (and $Q_{tot}$ as well). Therefore, it would be instructive to develop a simple model in the spirit of the coupled mode theory (CMT)\cite{Haus_book, Fan_03}. To begin, we would like to note that the in-plane hexapole mode profile, $|{\bf E}_{cav}({\bf r}_{||})|^2$, can be approximated in terms of the three $M$-point wavevectors of ${\bf k}_{M1}=(0,1)|{\bf k}_M|$, ${\bf k}_{M2}=(\sqrt{3}/2,1/2)|{\bf k}_M|$, and ${\bf k}_{M3}=(\sqrt{3}/2,-1/2)|{\bf k}_M|$ with $|{\bf k}_M| = 2 \pi / \sqrt{3} a$. For example, $|{\bf E}_{(cav)}({\bf r}_{||})|^2 \approx | \sin(\alpha{\bf k}_{M1} \cdot {\bf r}_{||}) - \sin(\alpha{\bf k}_{M2} \cdot {\bf r}_{||}) - \sin(\alpha{\bf k}_{M3} \cdot {\bf r}_{||}) |^2$ with a correction factor $\alpha > 1$\cite{Painter_03_PRB}. For our hexapole mode, $\alpha$ is $\sim1.1$ based on the ${\bf k}$-space intensity distribution, $|{\tilde{\bf E}}_{(cav)}({\bf k})|^2$\cite{Kim_PRB_06}. This is the reason the outskirt region of the hexapole mode resembles the $M$-point Bloch modes\cite{GHKim04}. 
  

\begin{figure}[t]
\centering\includegraphics[width=8.5cm]{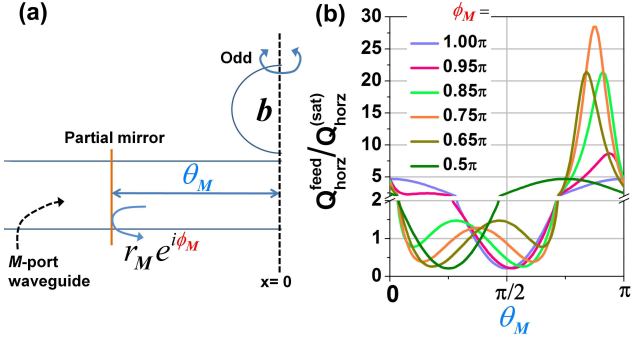}
\caption{\label{fig:figCMT} (a) In our coupled-mode model, the hexapole mode is assumed to couple only to the $\Gamma$-$M$ wavevectors. For further simplification, the solution for the entire system is {\em projected} into the $x$ direction. The hexapole symmetry ensures the odd mirror symmetry with respect to the $x = 0$ plane. (b) $Q_{horz}$ enhancement factors in Eq.~\ref{eq:eq4}. All plots assume $r_M$ = 0.6. }
\end{figure}

To simplify our CMT model, we assume that the hexapole mode couples {\em only} to the six $M$-point wavevectors, $\beta \times$(${\bf k}_{M1}$, ${\bf k}_{M2}$, ${\bf k}_{M3}$, $-{\bf k}_{M1}$, $-{\bf k}_{M2}$, $-{\bf k}_{M3}$), where $\beta$ is a scale factor $\sim 0.7$ as evidenced from the photonic band structure in Fig.~\ref{fig:fig1}(d). Now we emphasize that these $M$-point wavevectors ($\{ \beta {\bf k}_{M} \}$) are {\em closed} upon reflections at the six hexagonal facets. Thus, we may consider the whole set of wavevectors, $\{ \beta {\bf k}_{M} \}$, as a {\em channel} or a {\em port} in the CMT formulation [Fig.~\ref{fig:figCMT}(a)]. Under these assumptions, our hexapole mode can be viewed as {\em side-coupled} to the port\cite{sidecouple}. Then, the time evolution of the hexapole mode's energy amplitude ($b$) can be described by 

\begin{equation}
\frac{db}{dt} = \left[ i\omega_0 - \frac{\gamma_{vert} + \gamma_{horz}}{2} \right]b + \kappa_{M} S_{+M}, \label{eq:eq1}
\end{equation}
where the last term describes the {\em total incoming} power as a result of the feedback by the boundary termination. We assume that $S_{+M}$ is measured at the {\em inner} boundary of the partial mirror shown in Fig.~\ref{fig:figCMT}(a) and consists of the Bloch waves with $\{ \beta {\bf k}_{M} \}$. The CMT states that $\kappa_{M}$ cannot be arbitrarily determined but rather it should be connected by the decay rate into the waveguide port such that $\kappa_{M} =\sqrt{ 2\gamma_{horz} } e^{-i \theta_M}$.

When the effect of the photon feedback is weak as in the case of $p = \infty$, we can set $S_{+M} = 0$ in Eq.~\ref{eq:eq1}. Then, $|b|^2$ decays exponentially with $\gamma_{horz} = |\kappa_{M}|^2/2 = \omega_0 / Q^{(sat)}_{horz}$. However, when there is a feedback mechanism, $\gamma_{horz}$ is not a constant but varies with $\theta_M$ and $\phi_M$ (reflection phase). For this general situation, we need additional set of equations, which can be derived from the CMT\cite{ManolatouKh99}. For example, the outgoing power ($S_{-M}$) and the incoming power ($S_{+M}$) are related each other by $b$ through

\begin{equation}
S_{-M} = e^{-2 i \theta_{M}} S_{+M} + \sqrt{2 \gamma_{horz}} e^{-i \theta_{M}} b.\label{eq:eq3}
\end{equation}
We define the scattering matrix of the partial mirror such that $S_{+M} = r_{M} e^{i \phi_{M}} S_{-M}$. Then, the modified $Q_{horz}$ ($\equiv Q^{feed}_{horz}$) can be written using the definition of $Q$ ($\equiv \omega \times$ [total energy stored in the resonator]$/$[total power loss into the port]),

\begin{equation}
Q^{feed}_{horz} = \omega^{feed} \frac{|b|^2}{t^2_{M} |S_{-M}|^2}  \label{eq:eq32}
\end{equation}
In general, $\omega^{feed}$ (the resonance frequency in the presence of the feedback) differs from the original $\omega_0$ as we have seen from the FDTD result in Fig.~\ref{fig:fig3}(e). However, we can assume $\omega^{feed} \approx \omega_0$ since the fractional change in $\omega$ is much less than 1\%. Note that the partial mirror is assumed to be lossless such that $r^2_M + t^2_M =1$.

After solving $S_{-M}$ and $S_{+M}$ for $b$, we obtain the expression for the $Q_{horz}$ enhancement, $Q^{feed}_{horz}/Q^{(sat)}_{horz}$ ($= \gamma_{horz}/\gamma^{feed}_{horz}$), 

\begin{equation}
\frac{Q^{feed}_{horz}}{Q^{(sat)}_{horz}} = \frac{1+r^4_{M}-2r^2_{M} \cos (2 \phi_{M} + 4 \theta_{M})}{(1-r^2_{M})(1+r^2_{M} +2r_{M} \cos(\phi_{M} - 2\theta_{M}))}.\label{eq:eq4}
\end{equation}


Fig.~\ref{fig:figCMT}(b) shows plots of Eq.~\ref{eq:eq4} for several $\phi_M$ values, while $r_M$ is fixed at 0.6. The model also expects drastic modulations in $Q_{horz}$ depending where we locate the resonator with respect to the $|{\bf E}|^2$ envelop of the low-$Q$ Fabry-Perot type resonator. For example, when $\phi_M = 1.0\pi$, $Q_{horz}$ enhancement is maximized to be about 5 at $\theta_M = \pi$ (=effective half wavelength), which can be understood considering that $|{\bf E}|^2 = 0$ at $x = 0$. Interestingly, tuning $\phi_M$ slightly can greatly improve the enhancement; $\phi_M = 0.75\pi$ results in the maximum $Q^{feed}_{horz}/Q^{(sat)}_{horz}$ very close to 30. We also find that the $\phi_M$ tuning inevitably alters the optimal $\theta_M$ for $Q$, which agrees with the FDTD results in Fig.~\ref{fig:fig3}; the small tuning in $p$ changes the optimal $N$ $layer$ for $Q$. 

It should be noted that we have assumed only one ${\bf k}$-port and such a large enhancement holds true only for that port. 
However, if the defect state could excite many in-plane Bloch waves with different wavevectors thereby creating many ${\bf k}$-ports, the overall $Q_{horz}$ enhancement contributed from different ${\bf k}$-ports will be averaged out to be 1 as Eq.~\ref{eq:eq4} expects. For example, multiple ${\bf k}$ ports might be involved if $T$ were too thick (since this would allow more higher-order slab modes with many ${\bf k}$s) or if the boundary termination would not retain the hexapole symmetry (since ${\bf k}_{M}$ cannot be conserved upon reflections). 

In summary, we study a quasi-bound photon state within a PhC that does not posses a rigorous PBG. We show that the density of in-plane Bloch modes can be controlled by the termination of the finite-size PhC. The coupling strength between the bound state and the extended Bloch modes can be reduced by more than a factor of 40 in the case of a thick PhC slab. The air-suspended slab structure assumed in this study may appear impractical. However, low index cladding material can be placed underneath as a supporting structure or a metal cladding to the sides of the hexagonal boundary can be used a clamp\cite{J_Huang_10}, which could also be used as an electrode for current injection. By removing the thickness constraint of $T \approx \lambda / 2n_{eff}$, many unconventional cavity designs which were previously discarded because they cannot support a PBG can now be reconsidered, such as the use of a {\em thick}-slab for the design of current-injection nanolasers\cite{S_H_Kim_12}.

\begin{acknowledgments}
This work was supported by the Defense Advanced Research Projects Agency under the Nanoscale Architecture for Coherent Hyperoptical Sources program (W911NF-07-1-0277) and by the National Science Foundation under NSF CIAN ERC (EEC-0812072). A. Homyk appreciates the generous support of the ARCS Foundation.
\end{acknowledgments}


\bibliography{HighQ_woPBG}

\providecommand{\noopsort}[1]{}\providecommand{\singleletter}[1]{#1}%
\begin{thebibliography}{29}
\expandafter\ifx\csname natexlab\endcsname\relax\def\natexlab#1{#1}\fi
\expandafter\ifx\csname bibnamefont\endcsname\relax
  \def\bibnamefont#1{#1}\fi
\expandafter\ifx\csname bibfnamefont\endcsname\relax
  \def\bibfnamefont#1{#1}\fi
\expandafter\ifx\csname citenamefont\endcsname\relax
  \def\citenamefont#1{#1}\fi
\expandafter\ifx\csname url\endcsname\relax
  \def\url#1{\texttt{#1}}\fi
\expandafter\ifx\csname urlprefix\endcsname\relax\def\urlprefix{URL }\fi
\providecommand{\bibinfo}[2]{#2}
\providecommand{\eprint}[2][]{\url{#2}}

\bibitem[{\citenamefont{Purcell}(1946)}]{Purcell46}
\bibinfo{author}{\bibfnamefont{E.~M.} \bibnamefont{Purcell}},
  \bibinfo{journal}{Phys.\ Rev.} \textbf{\bibinfo{volume}{69}},
  \bibinfo{pages}{681} (\bibinfo{year}{1946}).

\bibitem[{\citenamefont{Yablonovitch}(1987)}]{Yablonovitch87}
\bibinfo{author}{\bibfnamefont{E.}~\bibnamefont{Yablonovitch}},
  \bibinfo{journal}{\prl} \textbf{\bibinfo{volume}{58}}, \bibinfo{pages}{2059}
  (\bibinfo{year}{1987}).

\bibitem[{\citenamefont{John}(1987)}]{Sajeev87}
\bibinfo{author}{\bibfnamefont{S.}~\bibnamefont{John}}, \bibinfo{journal}{\prl}
  \textbf{\bibinfo{volume}{58}}, \bibinfo{pages}{2486} (\bibinfo{year}{1987}).

\bibitem[{\citenamefont{Joannopoulos et~al.}(2008)\citenamefont{Joannopoulos,
  Johnson, Winn, and Meade}}]{Joannopoulos_book}
\bibinfo{author}{\bibfnamefont{J.~D.} \bibnamefont{Joannopoulos}},
  \bibinfo{author}{\bibfnamefont{S.~G.} \bibnamefont{Johnson}},
  \bibinfo{author}{\bibfnamefont{J.~N.} \bibnamefont{Winn}}, \bibnamefont{and}
  \bibinfo{author}{\bibfnamefont{R.~D.} \bibnamefont{Meade}},
  \emph{\bibinfo{title}{Photonic Crystals: Molding the Flow of Light}}
  (\bibinfo{publisher}{Princeton University Press},
  \bibinfo{address}{Princeton, NJ}, \bibinfo{year}{2008}),
  \bibinfo{edition}{2nd} ed.

\bibitem[{\citenamefont{Yablonovitch}(1993)}]{Yablonovitch_93}
\bibinfo{author}{\bibfnamefont{E.}~\bibnamefont{Yablonovitch}},
  \bibinfo{journal}{J. Opt. Soc. Am. B} \textbf{\bibinfo{volume}{10}},
  \bibinfo{pages}{283} (\bibinfo{year}{1993}).

\bibitem[{\citenamefont{Noda et~al.}(2000)\citenamefont{Noda, Tomoda, Yamamoto,
  and Chutinan}}]{Noda00}
\bibinfo{author}{\bibfnamefont{S.}~\bibnamefont{Noda}},
  \bibinfo{author}{\bibfnamefont{K.}~\bibnamefont{Tomoda}},
  \bibinfo{author}{\bibfnamefont{N.}~\bibnamefont{Yamamoto}}, \bibnamefont{and}
  \bibinfo{author}{\bibfnamefont{A.}~\bibnamefont{Chutinan}},
  \bibinfo{journal}{Science} \textbf{\bibinfo{volume}{289}},
  \bibinfo{pages}{604} (\bibinfo{year}{2000}).

\bibitem[{\citenamefont{Chow et~al.}(2000)\citenamefont{Chow, Lin, Johnson,
  Villeneuve, Joannopoulos, Wendt, Vawter, Zubrzycki, Hou, and
  Alleman}}]{Chow_Nature_00}
\bibinfo{author}{\bibfnamefont{E.}~\bibnamefont{Chow}},
  \bibinfo{author}{\bibfnamefont{S.}~\bibnamefont{Lin}},
  \bibinfo{author}{\bibfnamefont{S.}~\bibnamefont{Johnson}},
  \bibinfo{author}{\bibfnamefont{P.}~\bibnamefont{Villeneuve}},
  \bibinfo{author}{\bibfnamefont{J.}~\bibnamefont{Joannopoulos}},
  \bibinfo{author}{\bibfnamefont{J.}~\bibnamefont{Wendt}},
  \bibinfo{author}{\bibfnamefont{G.}~\bibnamefont{Vawter}},
  \bibinfo{author}{\bibfnamefont{W.}~\bibnamefont{Zubrzycki}},
  \bibinfo{author}{\bibfnamefont{H.}~\bibnamefont{Hou}}, \bibnamefont{and}
  \bibinfo{author}{\bibfnamefont{A.}~\bibnamefont{Alleman}},
  \bibinfo{journal}{Nature} \textbf{\bibinfo{volume}{407}},
  \bibinfo{pages}{983} (\bibinfo{year}{2000}).

\bibitem[{\citenamefont{Maldovan and Thomas}(2004)}]{Maldovan_NatMat}
\bibinfo{author}{\bibfnamefont{M.}~\bibnamefont{Maldovan}} \bibnamefont{and}
  \bibinfo{author}{\bibfnamefont{E.~L.} \bibnamefont{Thomas}},
  \bibinfo{journal}{Nat. Mater.} \textbf{\bibinfo{volume}{3}},
  \bibinfo{pages}{593} (\bibinfo{year}{2004}).

\bibitem[{\citenamefont{Yablonovitch et~al.}(1991)\citenamefont{Yablonovitch,
  Gmitter, Meade, Rappe, Brommer, and Joannopoulos}}]{YablonvitchGm91}
\bibinfo{author}{\bibfnamefont{E.}~\bibnamefont{Yablonovitch}},
  \bibinfo{author}{\bibfnamefont{T.~J.} \bibnamefont{Gmitter}},
  \bibinfo{author}{\bibfnamefont{R.~D.} \bibnamefont{Meade}},
  \bibinfo{author}{\bibfnamefont{A.~M.} \bibnamefont{Rappe}},
  \bibinfo{author}{\bibfnamefont{K.~D.} \bibnamefont{Brommer}},
  \bibnamefont{and} \bibinfo{author}{\bibfnamefont{J.~D.}
  \bibnamefont{Joannopoulos}}, \bibinfo{journal}{Physical Review Letters}
  \textbf{\bibinfo{volume}{67}}, \bibinfo{pages}{3380} (\bibinfo{year}{1991}).

\bibitem[{\citenamefont{Yoshie et~al.}(2004)\citenamefont{Yoshie, Scherer,
  Hendrickson, Khitrova, Gibbs, Rupper, Ell, Shchekin, and
  Deppe}}]{Yoshie_Nature_04}
\bibinfo{author}{\bibfnamefont{T.}~\bibnamefont{Yoshie}},
  \bibinfo{author}{\bibfnamefont{A.}~\bibnamefont{Scherer}},
  \bibinfo{author}{\bibfnamefont{J.}~\bibnamefont{Hendrickson}},
  \bibinfo{author}{\bibfnamefont{G.}~\bibnamefont{Khitrova}},
  \bibinfo{author}{\bibfnamefont{H.~M.} \bibnamefont{Gibbs}},
  \bibinfo{author}{\bibfnamefont{G.}~\bibnamefont{Rupper}},
  \bibinfo{author}{\bibfnamefont{C.}~\bibnamefont{Ell}},
  \bibinfo{author}{\bibfnamefont{O.~B.} \bibnamefont{Shchekin}},
  \bibnamefont{and} \bibinfo{author}{\bibfnamefont{D.~G.} \bibnamefont{Deppe}},
  \bibinfo{journal}{Nature} \textbf{\bibinfo{volume}{432}},
  \bibinfo{pages}{200} (\bibinfo{year}{2004}).

\bibitem[{\citenamefont{Noda et~al.}(2007)\citenamefont{Noda, Fujita, and
  Asano}}]{Noda_Nat_Photon_07}
\bibinfo{author}{\bibfnamefont{S.}~\bibnamefont{Noda}},
  \bibinfo{author}{\bibfnamefont{M.}~\bibnamefont{Fujita}}, \bibnamefont{and}
  \bibinfo{author}{\bibfnamefont{T.}~\bibnamefont{Asano}},
  \bibinfo{journal}{Nature} \textbf{\bibinfo{volume}{1}}, \bibinfo{pages}{449}
  (\bibinfo{year}{2007}).

\bibitem[{\citenamefont{Johnson et~al.}(1999)\citenamefont{Johnson, Fan,
  Villeneuve, Joannopoulos, and Kolodziejski}}]{JohnsonFa99}
\bibinfo{author}{\bibfnamefont{S.~G.} \bibnamefont{Johnson}},
  \bibinfo{author}{\bibfnamefont{S.}~\bibnamefont{Fan}},
  \bibinfo{author}{\bibfnamefont{P.~R.} \bibnamefont{Villeneuve}},
  \bibinfo{author}{\bibfnamefont{J.~D.} \bibnamefont{Joannopoulos}},
  \bibnamefont{and} \bibinfo{author}{\bibfnamefont{L.~A.}
  \bibnamefont{Kolodziejski}}, \bibinfo{journal}{Physical Review~B}
  \textbf{\bibinfo{volume}{60}}, \bibinfo{pages}{5751} (\bibinfo{year}{1999}).

\bibitem[{\citenamefont{Kim et~al.}(2012{\natexlab{a}})\citenamefont{Kim,
  Huang, and Scherer}}]{SHKim_OL2012}
\bibinfo{author}{\bibfnamefont{S.-H.} \bibnamefont{Kim}},
  \bibinfo{author}{\bibfnamefont{J.}~\bibnamefont{Huang}}, \bibnamefont{and}
  \bibinfo{author}{\bibfnamefont{A.}~\bibnamefont{Scherer}},
  \bibinfo{journal}{Opt. Lett.} \textbf{\bibinfo{volume}{37}},
  \bibinfo{pages}{488} (\bibinfo{year}{2012}{\natexlab{a}}).

\bibitem[{\citenamefont{Haus}(1984)}]{Haus_book}
\bibinfo{author}{\bibfnamefont{H.~A.} \bibnamefont{Haus}},
  \emph{\bibinfo{title}{Waves and Fields in Optoelectronics}}
  (\bibinfo{publisher}{Prentice-Hall}, \bibinfo{address}{Englewood Cliffs, N.
  J.}, \bibinfo{year}{1984}).

\bibitem[{\citenamefont{Madelung}(1978)}]{Madelung_book}
\bibinfo{author}{\bibfnamefont{O.}~\bibnamefont{Madelung}},
  \emph{\bibinfo{title}{Introduction to Solid-State Theory}}
  (\bibinfo{publisher}{Springer-Verlag}, \bibinfo{address}{Berlin},
  \bibinfo{year}{1978}), \bibinfo{note}{chap. 14}.

\bibitem[{\citenamefont{Srinivasan and Painter}(2002)}]{Srinivasan_02}
\bibinfo{author}{\bibfnamefont{K.}~\bibnamefont{Srinivasan}} \bibnamefont{and}
  \bibinfo{author}{\bibfnamefont{O.}~\bibnamefont{Painter}},
  \bibinfo{journal}{Opt. Express} \textbf{\bibinfo{volume}{10}},
  \bibinfo{pages}{670} (\bibinfo{year}{2002}).

\bibitem[{\citenamefont{Hinds}(1994)}]{Hinds}
\bibinfo{author}{\bibfnamefont{E.~A.} \bibnamefont{Hinds}},
  \emph{\bibinfo{title}{in Cavity Quantum Electrodynamics}}
  (\bibinfo{publisher}{Academic Press, Inc}, \bibinfo{address}{Orlando},
  \bibinfo{year}{1994}), \bibinfo{note}{edited by P. R. Berman}.

\bibitem[{\citenamefont{Ryu et~al.}(2003)\citenamefont{Ryu, Notomi, and
  Lee}}]{H_Y_Ryu_03}
\bibinfo{author}{\bibfnamefont{H.-Y.} \bibnamefont{Ryu}},
  \bibinfo{author}{\bibfnamefont{M.}~\bibnamefont{Notomi}}, \bibnamefont{and}
  \bibinfo{author}{\bibfnamefont{Y.-H.} \bibnamefont{Lee}},
  \bibinfo{journal}{Appl. Phys. Lett.} \textbf{\bibinfo{volume}{83}},
  \bibinfo{pages}{4294} (\bibinfo{year}{2003}).

\bibitem[{\citenamefont{Koshiba et~al.}(2001)\citenamefont{Koshiba, Tsuji, and
  Sasaki}}]{PML_PhC}
\bibinfo{author}{\bibfnamefont{M.}~\bibnamefont{Koshiba}},
  \bibinfo{author}{\bibfnamefont{Y.}~\bibnamefont{Tsuji}}, \bibnamefont{and}
  \bibinfo{author}{\bibfnamefont{S.}~\bibnamefont{Sasaki}},
  \bibinfo{journal}{Microwave and Wireless Components Letters, IEEE}
  \textbf{\bibinfo{volume}{11}}, \bibinfo{pages}{152 } (\bibinfo{year}{2001}).

\bibitem[{\citenamefont{Istrate et~al.}(2005)\citenamefont{Istrate, Green, and
  Sargent}}]{PhC_mirror_R}
\bibinfo{author}{\bibfnamefont{E.}~\bibnamefont{Istrate}},
  \bibinfo{author}{\bibfnamefont{A.~A.} \bibnamefont{Green}}, \bibnamefont{and}
  \bibinfo{author}{\bibfnamefont{E.~H.} \bibnamefont{Sargent}},
  \bibinfo{journal}{Phys. Rev. B} \textbf{\bibinfo{volume}{71}},
  \bibinfo{pages}{195122} (\bibinfo{year}{2005}).

\bibitem[{\citenamefont{Kim et~al.}(2006)\citenamefont{Kim, Kim, and
  Lee}}]{Kim_PRB_06}
\bibinfo{author}{\bibfnamefont{S.-H.} \bibnamefont{Kim}},
  \bibinfo{author}{\bibfnamefont{S.-K.} \bibnamefont{Kim}}, \bibnamefont{and}
  \bibinfo{author}{\bibfnamefont{Y.-H.} \bibnamefont{Lee}},
  \bibinfo{journal}{\prb} \textbf{\bibinfo{volume}{73}}, \bibinfo{eid}{235117}
  (pages~\bibinfo{numpages}{13}) (\bibinfo{year}{2006}).

\bibitem[{thi()}]{thin_slab_Q}
\bibinfo{note}{For the thin PhC slab, we find that $p \sim 1.0 a$ causes
  excessive scattering losses at the termination boundaries, resulting in the
  lower $Q$ than $p = \infty$ case. For fair comparions between {\em thin}- and
  {\em thick} slabs, we assume both modes oscillate at approximately same
  resonant wavelengths of 1300 nm, which can be achieved by choosing $a$ of the
  thin-slab and the thick-slab to be 410 nm and 350 nm, respectively.}

\bibitem[{\citenamefont{Fan et~al.}(2003)\citenamefont{Fan, Suh, and
  Joannopoulos}}]{Fan_03}
\bibinfo{author}{\bibfnamefont{S.}~\bibnamefont{Fan}},
  \bibinfo{author}{\bibfnamefont{W.}~\bibnamefont{Suh}}, \bibnamefont{and}
  \bibinfo{author}{\bibfnamefont{J.~D.} \bibnamefont{Joannopoulos}},
  \bibinfo{journal}{J. Opt. Soc. Am. A} \textbf{\bibinfo{volume}{20}},
  \bibinfo{pages}{569} (\bibinfo{year}{2003}).

\bibitem[{\citenamefont{Painter and Srinivasan}(2003)}]{Painter_03_PRB}
\bibinfo{author}{\bibfnamefont{O.}~\bibnamefont{Painter}} \bibnamefont{and}
  \bibinfo{author}{\bibfnamefont{K.}~\bibnamefont{Srinivasan}},
  \bibinfo{journal}{Phys. Rev. B} \textbf{\bibinfo{volume}{68}},
  \bibinfo{pages}{035110} (\bibinfo{year}{2003}).

\bibitem[{\citenamefont{Kim et~al.}(2004)\citenamefont{Kim, Lee, Shinya, and
  Notomi}}]{GHKim04}
\bibinfo{author}{\bibfnamefont{G.-H.} \bibnamefont{Kim}},
  \bibinfo{author}{\bibfnamefont{Y.-H.} \bibnamefont{Lee}},
  \bibinfo{author}{\bibfnamefont{A.}~\bibnamefont{Shinya}}, \bibnamefont{and}
  \bibinfo{author}{\bibfnamefont{M.}~\bibnamefont{Notomi}},
  \bibinfo{journal}{Opt. Express} \textbf{\bibinfo{volume}{12}},
  \bibinfo{pages}{6624} (\bibinfo{year}{2004}).

\bibitem[{sid()}]{sidecouple}
\bibinfo{note}{The side-coupling configuration may describe the situation
  better than the butt-coupling, because 1) the point-like defect structure
  occupies only a small fraction of the entire 2-D cavity and 2) the 2-D Bloch
  waves rather {\em freely} move from the left side to the right side without
  being assisted by the resonance of the defect.}

\bibitem[{\citenamefont{Manolatou et~al.}(1999)\citenamefont{Manolatou, Khan,
  Fan, Villeneuve, Haus, and Joannopoulos}}]{ManolatouKh99}
\bibinfo{author}{\bibfnamefont{C.}~\bibnamefont{Manolatou}},
  \bibinfo{author}{\bibfnamefont{M.~J.} \bibnamefont{Khan}},
  \bibinfo{author}{\bibfnamefont{S.}~\bibnamefont{Fan}},
  \bibinfo{author}{\bibfnamefont{P.~R.} \bibnamefont{Villeneuve}},
  \bibinfo{author}{\bibfnamefont{H.~A.} \bibnamefont{Haus}}, \bibnamefont{and}
  \bibinfo{author}{\bibfnamefont{J.~D.} \bibnamefont{Joannopoulos}},
  \bibinfo{journal}{IEEE Journal of Quantum Electronics}
  \textbf{\bibinfo{volume}{35}}, \bibinfo{pages}{1322} (\bibinfo{year}{1999}).

\bibitem[{\citenamefont{Huang et~al.}(2010)\citenamefont{Huang, Kim, and
  Scherer}}]{J_Huang_10}
\bibinfo{author}{\bibfnamefont{J.}~\bibnamefont{Huang}},
  \bibinfo{author}{\bibfnamefont{S.-H.} \bibnamefont{Kim}}, \bibnamefont{and}
  \bibinfo{author}{\bibfnamefont{A.}~\bibnamefont{Scherer}},
  \bibinfo{journal}{Opt. Express} \textbf{\bibinfo{volume}{18}},
  \bibinfo{pages}{19581} (\bibinfo{year}{2010}).

\bibitem[{\citenamefont{Kim et~al.}(2012{\natexlab{b}})\citenamefont{Kim,
  Huang, and Scherer}}]{S_H_Kim_12}
\bibinfo{author}{\bibfnamefont{S.-H.} \bibnamefont{Kim}},
  \bibinfo{author}{\bibfnamefont{J.}~\bibnamefont{Huang}}, \bibnamefont{and}
  \bibinfo{author}{\bibfnamefont{A.}~\bibnamefont{Scherer}},
  \bibinfo{journal}{J. Opt. Soc. Am. B} \textbf{\bibinfo{volume}{29}},
  \bibinfo{pages}{577} (\bibinfo{year}{2012}{\natexlab{b}}).

\end{thebibliography}

\end{document}